\documentclass[journal]{IEEEtran}
\usepackage{threeparttable}
\usepackage{cite}
\usepackage{mathrsfs}
\usepackage{bbm}
\usepackage{amssymb}
\usepackage{bbding}
\usepackage{threeparttable}
\usepackage{makecell}
\usepackage[mathcal]{euscript}
\usepackage{psfrag,calc,url,bm}
\usepackage{diagbox}
\usepackage{cite}
\usepackage{graphicx}
\usepackage{siunitx}
\usepackage{psfrag}
\usepackage{subfigure}
\usepackage{hyperref}
\usepackage{url}
\usepackage{stfloats}
\usepackage{amsmath}
\usepackage{float}
\usepackage{algorithmic}
\usepackage[ruled,linesnumbered,vlined]{algorithm2e}
\usepackage{color}
\usepackage{boxedminipage}
\usepackage{amsthm}
\usepackage{multirow}
\usepackage{setspace}
\usepackage{soul}
\usepackage{epstopdf}
\usepackage{booktabs}
\usepackage{cases}
\allowdisplaybreaks[3]

\allowdisplaybreaks[3]
\begin{document}

\title{GNN-Enabled Optimization of Placement and Transmission Design for UAV Communications}
\author{Qinyu Wang, Yang Lu,~\IEEEmembership{Member,~IEEE}, Wei Chen,~\IEEEmembership{Senior Member,~IEEE}, Bo Ai,~\IEEEmembership{Fellow,~IEEE}, \\ Zhangdui Zhong,~\IEEEmembership{Fellow,~IEEE}, and Dusit Niyato,~\IEEEmembership{Fellow,~IEEE}
\thanks{Qinyu Wang, Wei Chen, Bo Ai and Zhangdui Zhong are the School of Electronics and Information Engineering, Beijing Jiaotong University, Beijing 100044, China (e-mail: qinyuwang@bjtu.edu.cn; weich@bjtu.edu.cn; boai@bjtu.edu.cn; zhdzhong@bjtu.edu.cn).}
\thanks{Yang Lu is with the School of Computer Science and Technology, Beijing Jiaotong University, Beijing 100044, China (e-mail: yanglu@bjtu.edu.cn).}
\thanks{Dusit Niyato is with the School of Computer Science and Engineering, Nanyang Technological University, Singapore 639798 (e-mail: dniyato@ntu.edu.sg).}
}
\maketitle

\begin{abstract}
This paper applies graph neural networks (GNN) in UAV communications to optimize the placement and transmission design. We consider a multiple-user multiple-input-single-output UAV communication system where a UAV intends to find a placement to hover and serve users with maximum energy efficiency (EE). To facilitate the GNN-based learning, we adopt the hybrid maximum ratio transmission and zero forcing scheme to design the beamforming vectors and a feature augment is implemented by manually setting edge features. Furthermore, we propose a two-stage GNN-based model where the first stage and the second stage yield the placement and the transmission design, respectively. The two stages are connected via a residual and their learnable weights are jointly optimized by via unsupervised learning. Numerical results illustrate the effectiveness and validate the scalability to both UAV antennas and users of the proposed model. 
\end{abstract}

\section{Introduction}

Unmanned Aerial Vehicle (UAV) communications serve as an important complement to the terrestrial communications\cite{uav}. Due to flexible mobility and abundant line-of-sight links, the UAV can act as aerial base station or relay to  greatly enhance the wireless coverage of an interested area\cite{newuav}. Meanwhile, as an aerial vehicle, UAV can carry different scales of antennas, which allows them to meet quality of service and energy consumption requirements to adopt various application scenarios\cite{intro_survey}. So far, considerable efforts have been devoted to enhance UAV communications. In \cite{intro_SCA_EE}, a Dinkelbach method and successive convex approximation (SCA) based algorithm was employed to maximize the energy efficiency (EE) by optimizing communication scheduling, power allocation  and UAV trajectories. In \cite{place_beam}, an alternative optimization based algorithm involving weighted minimum mean square error method was proposed to optimize the placement and the MU-MIMO beamforming matrix of the UAV. Nevertheless, the mentioned algorithms may be time-costly due to their iterative frameworks, which are hard to realize real-time computation.

Applying artificial intelligence in UAV communication systems is an attractive and promising approach \cite{cv-channel}. In \cite{intro_LSTM}, an adaptive beam alignment scheme based on double-layer long short-term memory  model was proposed for intelligent reflecting surface assisted high-speed UAV communications to predict future positions and optimize beamforming vectors.  In \cite{intro_ddpg}, a deep deterministic policy gradient based algorithm was proposed for a UAV communication system with the aim of information rate maximization while satisfying the energy consumption requirements. The mentioned works are all based on traditional deep learning models. To leverage the graphical topology in wireless networks, graph neural networks (GNNs) draw increasing attention\cite{dl}. Especially, the GNN-based models was shown to be capability of handling the unseen problems such that they are scalable to different scenarios\cite{dl2}. The effectiveness of GNNs has been validated in typical wireless systems including MU-MISO \cite{new_dl3}, interference channels \cite{new_icnet} and physical-layer security \cite{new_hg}. Recently, some works have intended to apply GNNs to UAV communications. In \cite{intro_GNN_rate_angle},  a heterogeneous GNN named graph vision and communication model was proposed under deep reinforcement learning (DRL) framework to optimize UAV trajectories while improving the system throughput and fairness. In \cite{intro_GNN_DRL}, a GNN-assisted DRL algorithm was proposed, which took the input of UAV communication system states while yielding the actions to optimize the flight trajectory and scheduling strategy of the UAV to maximize the task offloading rate.

To the best of our knowledge, the GNN-enabled end-to-end learning for resource allocation in UAV communication systems has not been studied thus far. To fill this gap,  a joint placement and beamforming optimization problem is formulated and then, solved by a GNN-based model. Particularly, we formulate an EE maximization problem under constraints of the transmit power budget and the flyable area of the UAV. Based on the hybrid maximum ratio transmission (MRT) and zero forcing (ZF) scheme \cite{hyb}, we reformulate the considered problem to reduce the output dimension. Then, the UAV communication system is represented by a fully connected graph with defined node and edge features which are input into a proposed GNN-based model with multi-head attention and residual to map to the desired horizontal coordinate and transmission design of the UAV. Numerical results demonstrate the effectiveness of the proposed algorithm from perspectives of optimality, scalability and computational efficiency.

\section{Problem Definition}

\subsection{System Model and Problem Formulation}
\begin{figure}[t]
    \centering
\includegraphics[width=0.85\linewidth]{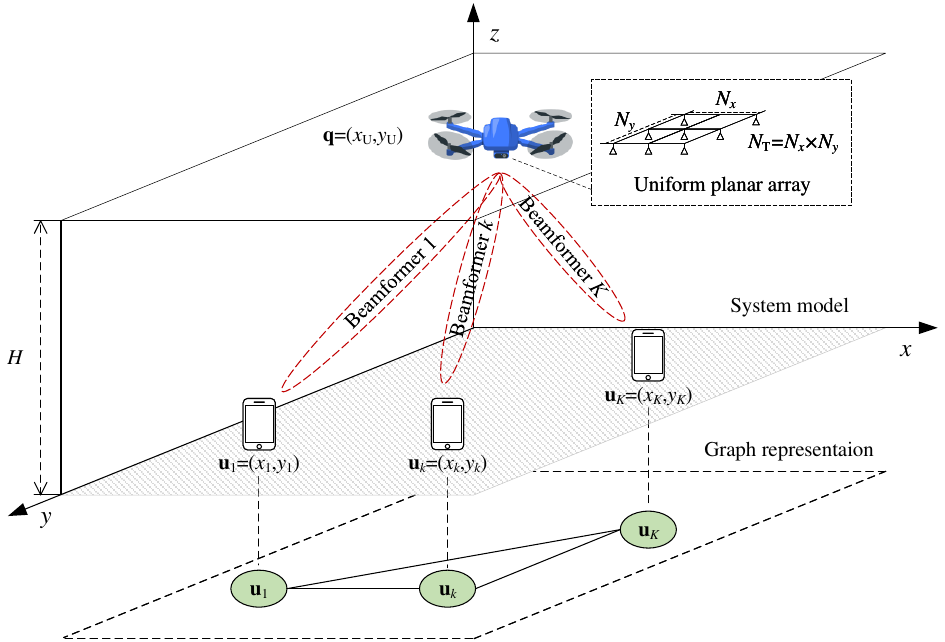}
    \caption{UAV communication system and graph representation.}
    \label{uavsys}
    \vspace{-0.5cm}
\end{figure}

Consider a multiple-input-single-output UAV communication system as shown in Figure \ref{uavsys}. The UAV is equipped with $N_{\mathrm T} = N_{\rm x} \times N_{\rm y}$  uniform planar array (UPA) antennas, and its flight altitude is fixed to $H$. The UAV aims to serve $K$ single-antenna ground users (GUs), and the horizontal coordinate of the $k$-th GU is denoted by ${{\bf u}_k}=(x_k,y_k)$.

Denote the horizontal coordinate of the UAV by ${\bf q}=(x_{\rm U},y_{\rm U})$. The channel between the UAV and the $k$-th GU is given by
\begin{flalign}
{\bm {\mathrm h}}_k\left({\bf q}\right) = \frac{\sqrt{\rho} \bm {\mathrm a}_k\left({\bf q}\right)}{\sqrt{{\Vert {\bf q}-{\bf u}_k \Vert}^2 + H^2}}\in {\mathbb C}^{N_{\rm T}},
\end{flalign}
where $\rho \triangleq (\lambda_{\rm c}/4\pi )^2$ with $\lambda_{\rm c}$ being the wavelength of carrier frequency, and $\bm {\mathrm a}_k({\bf q}) \in \mathbb{C}^{N_{\mathrm T}}$ denotes the steering vector from the UAV to the $k$-th GU, which is given by
\begin{flalign}	\label{channelmodel}	
\bm{{\mathrm a}}_k\left({\bf q}\right) =
& \left( 1, ...,  e^{-j\frac{2\pi b f_c}{c} \sin{\varpi_k}\left({\bf q}\right) (n_{\rm x}-1) \cos{\phi_k}\left({\bf q}\right)},..., \right.\\
&\left.  e^{-j\frac{2\pi b f_c}{c} \sin{\varpi_k}\left({\bf q}\right) (N_{\rm x}-1) \cos{\phi_k}\left({\bf q}\right)} \right)
\otimes  \nonumber\\
&\left( 1, ..., e^{-j\frac{2\pi b f_c}{c} \sin{\varpi_k}\left({\bf q}\right) (n_{\rm y}-1) \sin{\phi_k}\left({\bf q}\right)}, ...,\right.\nonumber\\
&\left.e^{-j\frac{2\pi b f_c}{c} \sin{\varpi_k}\left({\bf q}\right) (N_{\rm y}-1) \sin{\phi_k}\left({\bf q}\right)} \right),\nonumber
\end{flalign}
where $b$, $c$ and $f_c$ respectively denote the inter-antenna distance, the speed of light and the center frequency of carrier frequency. $n_{\rm x}$ and $n_{\rm y}$ denote the index of the UPA's row and column, respectively. $\varpi_k\left({\bf q}\right)$ denotes the vertical angle of departure (AoD). $\phi_k\left({\bf q}\right)$ denotes the horizontal AoD. They are respectively expressed as
\begin{subnumcases}
	{}
		\varpi_k\left({\bf q}\right) = \arcsin{\frac{H}{\sqrt{H^2+{\Vert {\bf q} - {\bf u}_k \Vert}^2}}}, \label{4a}\\
		 \phi_k\left({\bf q}\right) = \arccos{\frac{y_{\rm U} - y_{k}}{\Vert {\bf q}- {\bf u}_k \Vert}}.\label{4b}
\end{subnumcases}

The UAV serves all GUs over a common spectrum. The received signal at the $k$-th GU is expressed as 
\begin{flalign}
y_{k}\left({\bf q},\left\{{\bf w}_i\right\}\right) =  \sum\nolimits_{i=1}^K\bm{\mathrm h}_k^H\left({\bf q}\right){\bf w}_i s_i + n_k,	
\end{flalign}
\noindent
where $s_{i} \in \mathbb{C}$ represents the symbol for the $i$-th GU, which satisfies $\mathbb{E}\lbrace {|s_i|}^2 \rbrace=1$; ${\bf w}_k \in \mathbb{C}^{N_{\rm T} }$ denotes the corresponding beamforming vector; $n_{k}$ denotes AWGN of the $k$-th GU with power of $\sigma_{k}^2$. Then, the achievable data rate at the $k$-th GU is given by
\begin{flalign}
R_k \left({\bf q},\left\{{\bf w}_i\right\}\right)= \log_2\left(1+\frac{{\vert \bm {\mathrm h}_k^H\left({\bf q}\right) {\bf w}_k \vert}^2}{  \sum\nolimits_{i=1,i\ne k}^{K}{{\vert \bm {\mathrm h}_k^H\left({\bf q}\right) {\bf w}_j\vert}^2}+\sigma_{k}^2 }\right). \nonumber
\end{flalign}
Our goal is to maximize the energy efficiency subject to the transmit power budget constraint of the UAV. Then, the considered EE maximization problem is formulated as
\begin{subequations}\label{p0}
\begin{align}
& \mathop {\max }\limits_{{{\bf q},\left\{{{{\bf{w}}_i}}\right\}} }\mathop \frac{\sum\nolimits_{k=1}^K R_k \left({\bf q},\left\{{\bf w}_i\right\}\right)}{\sum\nolimits_{k=1}^K\left\| {{{\bf{w}}_k}} \right\|_2^2+P_{\rm c}} \label{p0:a}\\
{\rm s.t.}~&\sum\nolimits_{k=1}^K\left\| {{{\bf{w}}_k}} \right\|_2^2 \le {P_{\max}}, \label{p0:b}\\
& {\bf q} \in {\cal S},~\forall i,k \in \{1,2,...,K\},\label{p0:c}
\end{align}
\end{subequations}
where ${P_{\rm c}}$ denotes the constant circuit power due to information transmission, ${P_{\max}}$ denotes the transmit power budget, and ${\cal S}$ denotes the flyable area.

\subsection{Problem Reformulation}

Based on prior knowledge, we can simplify the mapping to enhance the learning capability. In particular, we simplify the output of the mapping. For the $k$-th  beamforming vector, it is composed of the power part denoted by $p_k$ and the direction part denoted by ${{\overline{\bf{w}}}_k}$, i.e.,
\begin{flalign}
{{\bf{w}}_k} = {\sqrt {p_k}}{{\overline{\bf{w}}}_k},~{\left\| {{{\overline {\bf{w}} }_k}} \right\|} = 1.
\end{flalign}

The MRT direction and the ZF direction are respectively given by
\begin{flalign}
{{{\overline {\bf{w}} }_k}}^{\left({\rm MRT}\right)} \left( {\bf q}\right) = \frac{{\bm {\mathrm h}}_k\left({\bf q}\right)}{\left\|{\bm {\mathrm h}}_k\left({\bf q}\right)\right\|}
\end{flalign}
and
\begin{flalign}
{{{\overline {\bf{w}} }_k}}^{\left({\rm ZF}\right)} \left( {\bf q}\right) = \frac{{\bm {\mathrm u}}_k\left({\bf q}\right)}{\left\|{\bm {\mathrm u}}_k\left({\bf q}\right)\right\|},
\end{flalign}
where ${\bf u}_k\left({\bf q}\right)$ is the $k$-th column of ${\bf U}({\bf q})\in{\mathbb C}^{N_{\rm T}\times K}$ given by \eqref{U_q} where ${\bf G}({\bf q})\triangleq [{\bf h}_{k}^H({\bf q}),...,{\bf h}_{k}^H({\bf q})]\in{\mathbb C}^{ K\times N_{\rm T}}$.
\begin{flalign}\label{U_q}
{\bf{U}}\left( {\bf{q}} \right) = 
{\bf{G}}{\left( {\bf{q}} \right)^H}{\left( {{\bf{G}}\left( {\bf{q}} \right){\bf{G}}{{\left( {\bf{q}} \right)}^H}} \right)^{ - 1}}, {N_{\rm{T}}} \ge K.
\end{flalign}

The hybrid MRT and ZF scheme is represented by the linear combination of the MRT direction and the ZF direction \cite{hyb}, which is given by
\begin{flalign}\label{hybrid}
{\overline {\bf w}}_k \left({\bf q},\alpha_k\right) = \frac{\alpha_k {{{\overline {\bf{w}} }_k}}^{\left({\rm MRT}\right)}\left({\bf q}\right) + \left(1-\alpha_k\right){{{\overline {\bf{w}} }_k}}^{\left({\rm ZF}\right)}\left({\bf q}\right)}{\left\|\alpha_k {{{\overline {\bf{w}} }_k}}^{\left({\rm MRT}\right)}\left({\bf q}\right) + \left(1-\alpha_k\right){{{\overline {\bf{w}} }_k}}^{\left({\rm ZF}\right)}\left({\bf q}\right)\right\|},
\end{flalign}
where $\alpha_k \in [0,1]$ denotes the hybrid coefficient.

With the hybrid MRT and ZF scheme, the problem \eqref{p0} is approximated by
\begin{subequations}\label{p1}
\begin{align}
& \mathop {\max }\limits_{{{\bf q},\left\{{{p_i},{\alpha_i}}\right\}} }\mathop \frac{\sum\nolimits_{k=1}^K R_k \left({\bf q},\left\{{{p_i},{\alpha_i}}\right\}\right)}{\sum\nolimits_{k=1}^K p_k + P_{\rm c}} \label{p1:a}\\
{\rm s.t.}~&\sum\nolimits_{k=1}^K p_k \le {P_{\max}}, \label{p1:b}\\
&\alpha_k\in \left(0,1\right],\label{p1:c}\\
& {\bf q} \in {\cal S},~\forall i,k \in \{1,2,...,K\},\label{p1:d}
\end{align}
\end{subequations}
where 
\begin{flalign}
R_k & \left({\bf q},\left\{{{p_i},{\alpha_i}}\right\}\right) =\nonumber\\
&\log_2\left(1+\frac{p_k{\vert \bm {\mathrm h}_k^H\left({\bf q}\right) {\overline {\bf w}}_k \left({\bf q},\alpha_k\right) \vert}^2}{  \sum\nolimits_{i=1,i\ne k}^{K}{p_i{\vert \bm {\mathrm h}_k^H\left({\bf q}\right) {\overline {\bf w}}_i \left({\bf q},\alpha_i\right)\vert}^2}+\sigma_{k}^2 }\right).\nonumber
\end{flalign}

\subsection{Graph Representation and Feature Augment}

In order to leverage the graph-structured topology information of wireless networks while handling the problem \eqref{p1}, we  convert the considered system into a fully connected directed graph denoted by $\mathcal{G} = (\mathcal{V},\mathcal{E})$. Here,  $\mathcal{V}$ denotes the set of nodes, and $\mathcal{E}$ denotes the set of edges. Each node represents a GU with node feature being its coordinate while the edge from the $i$-th node to the $j$-th node represents inter-GU impact (e.g., resource competition) with edge feature being $e_{i,j} \triangleq {\bf d}_{i,j} = {\bf u}_i-{\bf u}_j$. Then, the task of the GNN-enabled resource allocation is to construct and train a model to map the graph to a (near-)optimal solution to the problem \eqref{p1}. Mathematically, the model is given by 
\begin{flalign}\label{mapping}
    \left\{{\bf{q}},\left\{ {{p_i},{\alpha _i}} \right\}\right\} = {\Pi _{\bm\theta} } \left(\mathcal{G}\right),
\end{flalign}
where ${\bm\theta}$ denotes the trainable parameters.

\section{GNN-based Model Architecture}

\begin{figure}[t]
    \centering
\includegraphics[width=0.95\linewidth]{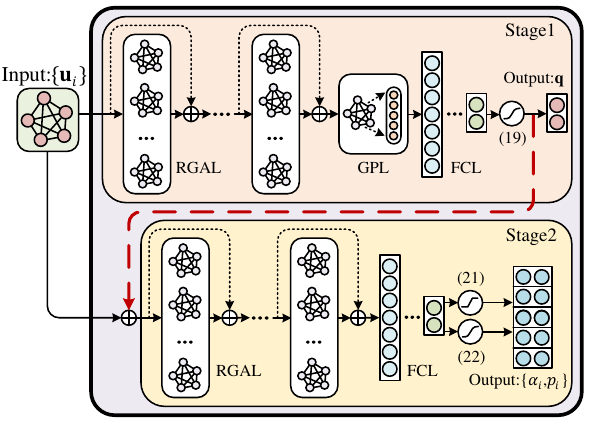}
    \caption{The architecture of the proposed two-stage GNN-based model.}
    \label{model}
\end{figure}

This section gives a realization of the model in \eqref{mapping} as illustrated in Figure \ref{model}. Particularly, the proposed GNN-based model is two-stage. The first stage generates a (rough) coordinate estimation of the UAV which is input into the second stage via a residual connection  together with the original graph to obtain the transmission design. The outputs of the two stages are both included in the loss function such that the placement and the transmission design of the UAV are jointly optimized instead of alternately optimized. At last, the proposed model is trained in an unsupervised manner.

The structure of the two stages are summarized here. 
\begin{enumerate}
    \item The first stage comprises $L_{1}$  residual graph attention layers (RGAL), a graph pooling layer (GPL) and $M_{1}$ fully connected layers (FCL). A customized Sigmod function guarantees that the output  horizontal coordinate of the UAV satisfies \eqref{p1:d}.  
    \item The second stage comprises $L_{2}$ RGALs and $M_{2}$ FCLs. A scale operation \cite{lugnn} and a customized Sigmod function guarantee that the output total power consumption and hybrid coefficients satisfy \eqref{p1:b} and \eqref{p1:c}, respectively.
\end{enumerate}

Different from the second stage with node-level readout, the first stage includes a GPL to facilitate the graph-level readout as the estimate of the coordinate of the UAV relies on all location information of GUs. Particularly, the GPL is given by 
\begin{flalign}
    {\bf q}^{(L_1+1)} = {\rm Pooling}\left(\{{\bf u}_{1}^{(L_{1})}, \ldots, {\bf u}_{K}^{(L_{1})}\}  \right),
\end{flalign}
where ${\bf q}^{(L_1+1)}$ denotes the pooling output which is input into the FCLs in the first stage and ${\bf u}_{k}^{(L_{1})}$ denotes the output associated with the $k$-th node of the $L_1$-th RGAL. 

The residual connection between the two stages is inspired by the fact that the transmission design depends on the coordinates of both UAV and GUs. With the residual connection, the node features  $\{{{\bf u}}_i\}$ are updated by $\{{\widehat {\bf u}}_i\}$, expressed as follows: 
\begin{equation}
 \mathbf{\widehat{u}}_{i} \triangleq \text{Com} ( {\bf u}_{i}, \bf q )
\end{equation}
where \text{Com}($\cdot$) represents the concatenation operation. For clarity, we denote the graph with $\{{\widehat {\bf u}}_i\}$ by $\widehat {\mathcal G}$, which is also the input of the second stage.

The RGALs in the first stage and the second stage share the same structure. Instead of vanilla message passing mechanism, the multi-head attention enabled aggregation is adopted to facilitate each node to focus on the relationship (i.e., the inter-GU impact in the first stage and the inter-GU interference in the second stage) from its neighbours based on their importance, and the residual-based update is adopted to mitigate the over-smoothing issue. Similarly, the FCLs in the first stage and the second stage share the same structure, which are utilized to re-shape the extracted features into the required dimensions. In the following, we give the detailed processes of RGAL and FCL for the two stages.

\subsection{Residual Graph Attention Layer}

Each RGAL includes two main ingredients, i.e., the multi-head attention enabled aggregation and the residual-based update.  

\subsubsection{Multi-head attention enabled aggregation}
Each node learns the graph-level information by aggregating node features of its neighboring node as well as the associated edge features. To distinguish the importance of each neighboring node, the multi-head attention mechanism is adopted for each node to assign attention weights for it neighboring nodes.  

For the $l$-th RGAL ($l \in {\mathcal L} \triangleq \{1,...,L_{1}\}$ or $\{1,,...,L_{2}\}$), denote the input node feature of the $i$-th node by $\overline{\bf u}^{(l)}_{i} \in \mathbb{R}^{G(l)}$, where $G(l)$ denotes the dimension. $D(l)$-head self-attention mechanisms are employed, and the attention weight of the node pair  $i\rightarrow j$ under the $d$-th ($d \in \{1,...D(l)\}$) attention head is given by \eqref{attention_coefficient}, where 
$\mathbf{a}^{(l)}_{d} \in \mathbb{R}^{\widetilde G(l)}$ denote the learnable attention vector, $\mathbf{\Theta}^{(l)}_{s,d} \in \mathbb{R}^{\widetilde G(l) \times G(l)} $, $\mathbf{\Theta}^{(l)}_{t,d} \in \mathbb{R}^{\widetilde G(l) \times G(l)} $ and $\mathbf{\Theta}^{(l)}_{e,d} \in \mathbb{R}^{\widetilde G(l) \times G(l)} $ denote the learnable linear transformation matrices for the source node feature, target node feature, and edge feature, respectively, and $\widetilde G(l)$ denotes the transformed dimension.  Then, the aggregated feature of the $i$-th node under the $d$-th attention head is given by 
\begin{flalign}
\label{aggregation}
\beta_{i,d}^{(l)} = \alpha^{(l)}_{i,i}\mathbf{\Theta}^{(l)}_{s,d} {\bf \overline{u}}^{(l)}_{i} + \sum_{j \in \mathcal{N}(i)} \alpha^{(l)}_{i,j}\mathbf{\Theta}^{(l)}_{t,d}{\bf \overline{u}}^{(l)}_{j}  \in {\mathbb R}^{\widetilde G(l)}.
\end{flalign}

\begin{figure*}[!t]
\centering
\begin{equation}
\alpha_{i,j,d}^{(l)} = \frac{\exp \left( \mathrm{LeakyReLU} \left( \mathbf{\Theta}^{(l)}_{s,d} \mathbf{\overline{u}}^{(l)}_{i} + \mathbf{\Theta}^{(l)}_{t,d} \mathbf{\overline{u}}^{(l)}_{j}  + \mathbf{\Theta}^{(l)}_{e,d} \mathbf{d}^{(l)}_{i,j}  \right)^T \mathbf{a}^{(l)}_{d} \right)}
{\sum_{k \in \mathcal{N}(i) \cup \{ i \}} \exp \left( \mathrm{LeakyReLU} \left( 
\mathbf{\Theta}^{(l)}_{s,d} \mathbf{\overline{u}}^{(l)}_{i} + \mathbf{\Theta}^{(l)}_{t,d} \mathbf{\overline{u}}^{(l)}_{k}  + \mathbf{\Theta}^{(l)}_{e,d} \mathbf{d}^{(l)}_{i,k}  \right)^T \mathbf{a}^{(l)}_{d} \right)}
\label{attention_coefficient}
\end{equation}
\hrule
\end{figure*}

\subsubsection{Residual-based update}
Each node updates its node features based on its aggregated information. To mitigate the over-smoothing issue by making the updated node features of nodes distinguishable, the input node feature is also included via a jump residual connection.

Denote the output node feature of the $i$-th node of the $l$-th RGAL by  $\widetilde{\bf u}^{(l)}_{i} \in \mathbb{R}^{\widetilde G(l) \times D(l)}$. The $i$-th node is updated through the $l$-th RGAL by
\begin{flalign}
\widetilde{\bf u}^{(l)}_{i} = &\text {ReLU} \left( \text{Com}\left(\left\{ \beta_{d,i}^{(l)} \right\}_{d=1}^{D^{(l)}}\right) + \mathbf{\overline \Theta}^{(l)} {\bf \overline{u}}^{(l)}_{i} \right) \\
&\in {\mathbb R}^{\widetilde G(l) \times D(l)},\nonumber
\end{flalign}
where $\mathbf{\overline \Theta}^{(l)} \in {\mathbb R}^{ (\widetilde G(l) \times D(l)) \times G(l)}$ denote the trainable parameters of a feed-forward NN due to the jump residual connection, and the ${\rm ReLU}(\cdot)$ denotes the ReLU function.

\subsection{Fully Connected Layer and Activation Function}

The FCL is to ``decode" the extracted features by the RGALs to the desired solution, i.e., $\bf q$ or $\{\alpha_i,p_i\}$. 

In the first stage, $M_1$ FCLs are attached to the GPL to carry out graph-level computation, among which the $m$-th ($m \in \{1,...,M_{1}\}$) FCL is given by
\begin{flalign}
\widetilde{\bf u}^{(m)} = \text{ReLU} \left( \text{BN} \left( \mathbf{\widetilde \Theta}^{(m)} \overline{\bf u}^{(m)} \right) \right) \in {\mathbb R}^{\widetilde F(m)},
\end{flalign}
where $\widetilde{\bf u}^{(m)}$, $\overline{\bf u}^{(m)}$, $\mathbf{\widetilde \Theta}^{(m)}$ and ${{\widetilde F}(m)}$ denote the graph-level input and out feature, learnable linear transformation matrices of  and transformed dimension, respectively, and ${\rm BN}(\cdot)$ denotes the batch normalization function. Particularly, ${{\widetilde F}(M_1)}=2$ and the output of the $M_1$-th FCL is input into the following activation function according to the constraint \eqref{p1:d}:
\begin{flalign}\label{af_q}
{\mathbf{q}} = \left( N_{\rm x} \times \delta\left(\widetilde{\bf u}^{(M_1)}[1]\right), N_{\rm y} \times \delta\left(\widetilde{\bf u}^{(M_1)}[2]\right) \right),
\end{flalign}
where $\delta(\cdot): {\mathbb R}\rightarrow (0,1)$ denotes the Sigmod function. 

In the second stage, $M_2$ FCLs accept the input of the updated node feature, e.g., ${\widetilde {\bf u}}_i^{(L_2)}$, and allow parameter sharing among nodes. The $m$-th ($m \in \{1,...,M_{2}\}$) FCL is given by
\begin{flalign}
\widetilde{\bf u}^{(m)}_{i} = \text{ReLU} \left( \text{BN} \left( \mathbf{\widetilde \Theta}^{(m)} \overline{\bf u}^{(m)}_{i} \right) \right) \in {\mathbb R}^{\widetilde F(m)}.
\end{flalign}
Particularly, ${{\widetilde F}(M_2)}=2$ and the output of the $M_2$-th FCL is input into the following activation functions according to the constraints \eqref{p1:b} and \eqref{p1:c}:
\begin{flalign}\label{af_p}
&{p}_k = \left\{ \begin{array}{l}
\widetilde{\bf u}_k^{(M_2)}[1],~\sum\limits_{i=1}^K \widetilde{\bf u}_i^{(M_2)}[1] \le {P_{\max}},\\
\frac{\widetilde{\bf u}_k^{(M_2)}[1]}{{\sum\limits_{i = 1}^K \widetilde{\bf u}_i^{(M_2)}[1] }}{P_{\max }},~\sum\limits_{i=1}^K \widetilde{\bf u}_i^{(M_2)}[1]> {P_{\max}},
\end{array} \right.\\
\label{af_alpha}
&{\rm and}~{\alpha}_k = \delta\left( \widetilde{\bf u}_k^{(M_2)}[2]\right).
\end{flalign}

\subsection{Loss Function and Unsupervised Learning}

We use the objective function of the problem \eqref{p1} to construct the loss function as the output of FCLs is feasible to the problem \eqref{p1}. Besides, the unsupervised learning is adopted to train the proposed model. Therefore, the loss function of a batch of samples is given by 
\begin{flalign}\label{loss function}
{\cal  L}_{N}({\bm\theta} ) = \left.-\frac{1}{N} \sum\limits_{n=1}^N \left( \frac{\sum\limits_{k=1}^K R_k \left(\left\{\widehat{\bf{q}},\left\{ {\widehat{p}_i,\widehat{\alpha}_i} \right\}\right\}\right)}{\sum\limits_{k=1}^K \widehat{p}_k + P_{\rm c}}    
 \right) ^{(n)} \right| _{\bm\theta}, 
\end{flalign}
where ${\bm\theta} \triangleq \{ \mathbf{a}^{(l)}, \mathbf{\Theta}^{(l)}_{s}$, $\mathbf{\Theta}^{(l)}_{t}$, $\mathbf{\Theta}^{(l)}_{e}$, $\mathbf{\overline \Theta}^{(l)}$, $\mathbf{\widetilde \Theta}^{(m)} \}$ represents all trainable parameters and $N$ denotes the batch size.

\section{Numerical Results}
The section provides the numerical results to show the effectiveness of the proposed GNN-based model.

{\bf Simulation scenario.} 
The UAV is equipped with $N_{\mathrm T}\in\{2\times 3, 2\times4, 2\times5, 3\times 4, 2\times7 \}$ antennas at an altitude of $H=100$ m. The number of GUs is set as $K\in\{3,4,5,6,7\}$, and the GUs are randomly distributed within an area of $40\times40$ m$^{2}$. The channel model follows \eqref{channelmodel}. The noise power is set as $-80$ dBm. The center frequency, inter-antenna distance, transmit power budget and constant circuit power of the UAV are set as $2.4$ GHz, $0.0625$ m, $2.5N_{\mathrm T}$ W and $0.25 N_{\mathrm T}$ W, respectively \cite{UAV_parameter}. 

{\bf Dataset.} Five training sets with $(N_{\rm T},K)=\{(2\times4,4), (3\times4,4), (2\times3,6), (2\times5,6), (3\times4,6)\}$ are prepared and each training set includes $8,000$ unlabeled samples. For each training set, a $1,000$-sample validation set is prepared with the same setting. Besides, twelve test sets are prepared and each test set includes $1,000$ samples.

{\bf Implementation Details.} The learning rate is initialized as $4 \times 10^{-5}$. The adaptive moment estimation \cite{adam} is adopted as the optimizer during the training phase. The batch size is set as $32$ for $120$ epochs of training. Our implementation is developed using Python 3.10.14 with Pytorch 2.0.0 on a computer with AMD EPYC 9654 CPU and one RTX 4090 GPU (24 GB of memory).

\begin{figure}[t]
    \centering
\includegraphics[width=1\linewidth]{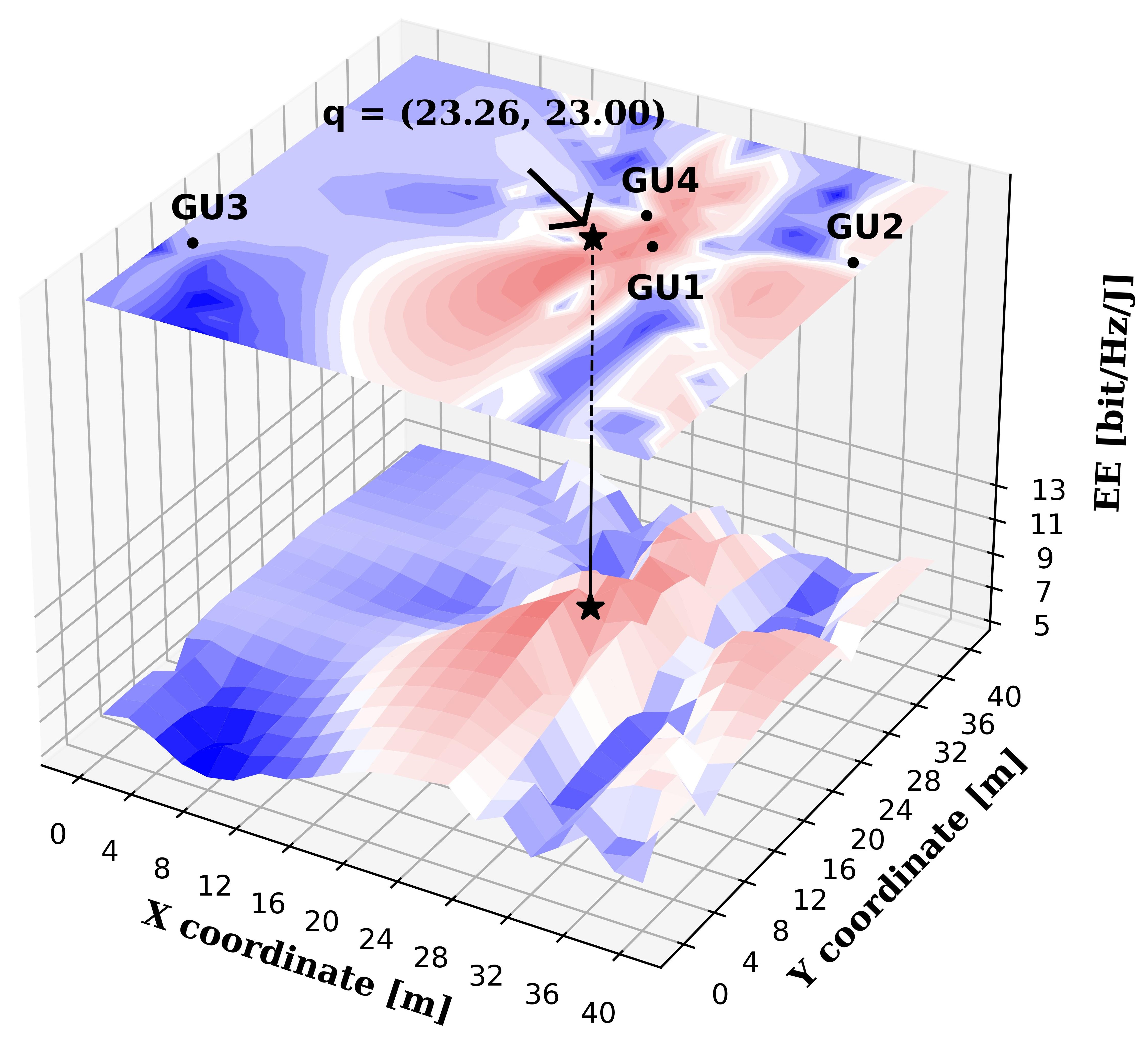}
    \caption{Illustration of optimization of placement and transmission design. }
    \label{example}
\end{figure}

Figure \ref{example} illustrate the effectiveness of the proposed model where the maximum EE of each position is calculated by an SCA-based algorithm. It is observed that the GNN-model is able to yield high-quality position and transmission design in this example. Besides, it takes several hours to exhaustively search to find the optimal position while it takes less than $1$ milliseconds by the proposed model.

Table \ref{generalization} evaluates the scalability of the proposed model and a multi-layer perceptron (MLP) is also trained as a baseline\footnote{We extend the MLP-based transmit desgin in \cite{new_MLP} to solve the considered problem as a baseline.}. It is observed that the proposed GNN-based model is scalable to both number of UAV antennas and GUs which are unseen in the training set, while the traditional MLP may not work. Note that the total power budget and constant circuit power are also dependent on the number of antennas. The scalability to these system parameters is also validated. Besides, the GNN-based model outperforms the MLP in terms of the EE due to more powerful feature extracting capability. 

\begin{table}
    \centering
    \begin{threeparttable}
    \caption{Scalability of GNN to different numbers of antennas and GUs in terms of EE [bit/Hz/J].}
    \begin{tabular}{c|c|c|c|c|c}
        \hline    
        $K^{({\rm Tr})}$ & $N_{\rm T}^{({\rm Tr})}$ & $K^{({\rm Te})}$ & $N_{\rm T}^{({\rm Te})}$ & MLP & GNN \\
        \hline
        \hline    
        $4$ & $3 \times 4$ & $4$ & $3 \times 4$ & $9.05$ & $9.46$ \\
        \hline
        \multirow{5}{*}{$4$} & \multirow{5}{*}{$2 \times 4$} & \multirow{3}{*}{$4$} & $2 \times 3$ & $10.81$ & $11.38$ \\
        \cline{4-6}
         &  &  & $2 \times 5$ & $9.60$ & $10.20$ \\
        \cline{4-6}
         &  &  & \multirow{3}{*}{$2 \times 4$} & $10.34$ & $10.95$ \\
        \cline{3-3} \cline{5-6}
         &  & $3$ &  & $\times$ & $10.13$ \\
        \cline{3-3} \cline{5-6}
         &  & $5$ &  & $\times$ & $10.46$ \\
         \hline
        \multirow{5}{*}{$6$} & \multirow{5}{*}{$3 \times 4$} & \multirow{3}{*}{$6$} & $2 \times 5$ & $9.07$ & $9.68$ \\
        \cline{4-6}
         &  &  & $2 \times 7$ & $9.13$ & $9.74$ \\
        \cline{4-6}
         &  &  & \multirow{3}{*}{$3 \times 4$} & $9.16$ & $9.76$ \\
        \cline{3-3} \cline{5-6}
         &  & $5$ &  & $\times$ & $10.00$ \\
        \cline{3-3} \cline{5-6}
         &  & $7$ &  & $\times$ & $8.98$ \\
        \hline
        \hline    
        \multicolumn{4}{c|}{Inference time}  & $0.40$ ms & $0.46$ ms \\
        \hline     
        \end{tabular}
        \begin{tablenotes}[flushleft]
            \small
            \item $K^{({\rm Tr})}/K^{({\rm Te})}$: Number of GUs in the training/test set.
            \item $N_{\rm T}^{({\rm Tr})}/N_{\rm T}^{({\rm Te})}$: Number of antennas in the training/test set.
        \end{tablenotes}
    \label{generalization}
    \end{threeparttable}
\end{table}

\begin{table}[t]
\centering
\caption{Ablation experiment in terms of EE [bit/Hz/J] with $K^{({\rm Tr})}=K^{({\rm Te})}=6$.}
\begin{tabular}{c c |c|c|c | c}
\hline
\multirow{2}*{FA}  &  \multirow{2}*{RC}  & \multicolumn{3}{c|}{$N_{\rm T}^{({\rm Tr})}=N_{\rm T}^{({\rm Te})}$} & \multirow{2}*{Inference time}\\
 \cline{3-5}
 &  &  $2\times 3$  & $2\times 5$  & $3\times 4$  &  \\
 \hline
 \hline
 $\times$ & $\checkmark$ & 6.46  & 9.44 & 9.73 & 0.45 ms\\
 \cline{3-6}
$\checkmark$  & $\times$ &  6.51 & 9.50 & 9.64 & 0.46 ms\\
 \cline{3-6}
$\checkmark$  & $\checkmark$ & 6.63 & 9.53 & 9.76 & 0.46 ms\\ 
 \hline
\end{tabular}
\label{table:ablation}
\begin{tablenotes}
        \footnotesize
        \item FA/RC: Feature argument/residual connection.
\end{tablenotes}
\end{table}

Table \ref{table:ablation} provides the ablation experiment to demonstrate the effectiveness of the feature argument (in Section II-C) and residual connection from the first stage to the second stage (in Section III). It is observed that both two mechanisms improve the obtained EE and have negligible impact on the inference time. Note that the two mechanisms can be extended to other GNN-based models for wireless resource allocation. 

\section{Conclusion}
This paper proposed a two-stage GNN-based model for UAV communications which yielded the placement in the first stage and the transmission design in the second stage. The proposed model was utilized to solve an EE maximization problem subject to power budget and flyable area via an unsupervised manner and shown to be with good scalability to both UAV antennas and GUs.


\begin{thebibliography}{1}

\bibitem{uav}
W. Mao et al., ``Energy consumption minimization in secure multi-antenna UAV-assisted MEC networks with channel uncertainty," \emph{IEEE Trans. Wireless Commun.}, vol. 22, no. 11, pp. 7185-7200, Nov. 2023.

\bibitem{newuav}
W. Mao et al., ``UAV-assisted communications in SAGIN-ISAC: Mobile user tracking and robust beamforming," early accessed by \emph{IEEE J. Sel. Areas Commun.}, 2024.

\bibitem{intro_survey}
Y. Liu, K. Xiong, Y. Lu, Q. Ni, P. Fan, and K. B. Letaief, ``UAV-aided wireless power transfer and data collection in Rician fading," \emph{IEEE J. Sel. Areas Commun.}, vol. 39, no. 10, pp. 3097-3113, Oct. 2021.

\bibitem{intro_SCA_EE}
X. Liu, Z. Liu, and M. Zhou, ``Fair energy-efficient resource optimization for green multi-NOMA-UAV assisted Internet of Things," \emph{IEEE Trans. Green Commun. Networking}, vol. 7, no. 2, pp. 904-915, June 2023.

\bibitem{place_beam}
C. Zeng, J.-B. Wang, M. Xiao, Y, Pan, Y. Chen, H. Yu, and J. Wang, ``Generalized optimization method of placement and beamforming design for multiuser MIMO UAV communications," \emph{IEEE Wireless Commun. Lett.}, vol. 13, no. 2, pp. 402-406, Feb. 2024.

\bibitem{cv-channel}
Z. Hua, Y. Lu, G. Pan, K. Gao, D. B. d. Costa, and S. Chen, ``Computer vision-aided mmWave UAV communication systems," \emph{IEEE Internet Things J.}, vol. 10, no. 14, pp. 12548-12561, July 2023.

\bibitem{intro_LSTM}
Q. Deng, X. Chen, X. Liang, F. Shu, J. Du, G. Yu, and J. Wang, ``Adaptive beam alignment and optimization for IRS-aided high-speed UAV communications," \emph{IEEE Trans. Green Commun. Networking}, vol. 7, no. 3, pp. 1583-1595, Sept. 2023.

\bibitem{intro_ddpg}
Y. Yu, J. Tang, J. Huang, X. Zhang, D. K. C. So, and K.-K. Wong, ``Multi-objective optimization for UAV-assisted wireless powered IoT networks based on extended DDPG algorithm," \emph{IEEE Trans. Commun.}, vol. 69, no. 9, pp. 6361-6374, Sept. 2021.

\bibitem{dl}
Y. Li, Y. Lu, R. Zhang, B. Ai, and Z. Zhong, ``Deep learning for energy efficient beamforming in MU-MISO networks: A GAT-based approach," \emph{IEEE Wireless Commun. Lett.}, vol. 12, no. 7, pp. 1264-1268, July 2023.

\bibitem{dl2}
Y. Li, Y. Lu, B. Ai, O. A. Dobre, Z. Ding and D. Niyato, ``GNN-based beamforming for sum-rate maximization in MU-MISO networks," \emph{IEEE Trans. Wireless Commun.}, vol. 23, no. 8, pp. 9251-9264, Aug. 2024.

\bibitem{new_dl3}
Y. Li, Y. Lu, B. Ai, Z. Zhong, D. Niyato, and Z. Ding, ``GNN-enabled max-min fair beamforming," \emph{IEEE Trans. Veh. Technol.}, vol. 73, no. 8, pp. 12184-12188, Aug 2024.

\bibitem{new_icnet}
C. He, Y. Li, Y. Lu, B. Ai, Z. Ding, and D. Niyato, ``ICNet: GNN-enabled beamforming for MISO interference channels with statistical CSI," \emph{IEEE Trans. Veh. Technol.}, vol. 73, no. 8, pp. 12225-12230, Aug 2024.

\bibitem{new_hg}
Z. Song, Y. Lu, X. Chen, B. Ai, Z. Zhong, and D. Niyato, ``A deep learning framework for physical-layer secure beamforming," early accessed in \emph{IEEE Trans. Veh. Technol.}, 2024.

\bibitem{intro_GNN_rate_angle}
X. Zhang, H. Zhao, J. Wei, C. Yan, J. Xiong, and X. Liu, ``Cooperative trajectory design of multiple UAV base stations with heterogeneous graph neural networks," \emph{IEEE Trans. Wireless Commun.}, vol. 22, no. 3, pp. 1495-1509, March 2023.

\bibitem{intro_GNN_DRL}
K. Li, W. Ni, X. Yuan, A. Noor, and A. Jamalipour, ``Exploring graph
 neural networks for joint cruise control and task offloading in UAV
enabled mobile edge computing," in \emph{Proc. VTC-Spring}, pp. 1-6, 2023.

\bibitem{hyb}
W. Guo, Y. Lu, H. Du, B. Ai, D. Niyato, and Z. Ding, ``Hybrid MRT and ZF learning for energy-efficient transmission in multi-RIS-assisted networks," \emph{IEEE Trans. Veh. Technol.}, vol. 73, no. 8, pp. 12247-12251, Aug. 2024.

\bibitem{lugnn}
Y. Lu et al., ``Graph neural networks for wireless networks: Graph representation, architecture and evaluation," early accessed by 
 \emph{IEEE Wireless Commun.}, 2024.

\bibitem{UAV_parameter}
D. Xu, Y. Sun, D. W. K. Ng, and R. Schober, “Multiuser MISO UAV communications in uncertain environments with no-fly zones: Robust trajectory and resource allocation design,” \emph{IEEE Trans. Commun.}, vol. 68, no. 5, pp. 3153-3172, May 2020.

\bibitem{adam}
D. P. Kingma and J. Ba, ``Adam: A method for stochastic optimization," arXiv:1412.6980.




\bibitem{new_MLP}
C. Hu, Y. Lu, H. Du, M. Yang, B. Ai, and D. Niyato, ``AI-empowered RIS-assisted networks: CV-enabled RIS selection and DNN-enabled transmission," early accessed in \emph{IEEE Trans. Veh. Technol.}, 2024.


\end{thebibliography}
\end{document}